\documentclass[12pt,draftcls,onecolumn,peerreview]{IEEEtran}
\usepackage{cite,graphicx,subfigure,amsmath,amssymb,array,verbatim,longtable,setspace,tikz,helvet}
\usepackage{multirow}
\usepackage{hhline}
\usepackage{lscape}
\usepackage{caption} 
\begin{document}

\title{Cooperative Network Coded ARQ Strategies for Two Way Relay Channel}

\author{Rasit~Tutgun and Emre~Aktas,~\IEEEmembership{Senior Member,~IEEE}
 \thanks{R. Tutgun is with TUBITAK Uzay Space Technologies Research Institute, Satellite 
 Technologies Department, Communication Systems Group, Ankara, 06531 TURKEY.
 e-mail: rasit.tutgun@tubitak.gov.tr}%
 \thanks{E. Aktas is with the Department
 of Electrical and Electronics Engineering, Hacettepe University,
 Beytepe, Ankara, 06800 TURKEY.
 e-mail: aktas@ee.hacettepe.edu.tr.}%
}

\maketitle

\begin{abstract}
In this paper, novel cooperative automatic repeat request (ARQ) methods with network coding are proposed for two way relaying network.
Upon a failed transmission of a packet, the network enters cooperation phase, where the retransmission of the packets is aided by the relay node.
The proposed approach integrates network coding into cooperative ARQ, 
aiming to improve the network throughput by reducing the number of 
retransmissions.
For successive retransmission, three different methods for choosing 
the retransmitting node are considered. 
The throughput of the methods are analyzed and compared. The analysis 
is based on binary Markov channel which 
takes the correlation of the channel coefficients in time into account. 
Analytical results
show that the proposed use of network coding result in throughput performance superior to traditional ARQ and cooperative ARQ without network coding.
It is also observed that correlation can have significant effect 
on the performance of the proposed cooperative network coded ARQ approach. 
In particular the proposed approach is advantageous 
for slow to moderately fast fading channels.
\end{abstract}

\begin{IEEEkeywords}
cooperative communication, automatic repeat request, network coding
\end{IEEEkeywords}


\section{Introduction}
\IEEEPARstart{C}{ooperative} communication has become an active research topic 
due to its ability to 
benefit from spatial diversity with the help of cooperative nodes (relays). 
The main goal of cooperation is to achieve diversity gain by 
using statistically independent channels for transmission 
\cite{meu71,cg79,sea03,ltw04,nhh04}. 
Most of these studies reveal the benefits of 
cooperative communication on the design and performance of physical layer. 
Recently, there have been other 
studies which investigate the advantages of cooperative methods on the 
higher layer methods such as automatic repeat request (ARQ).

ARQ is an error control mechanism for increasing reliability in modern communication systems.
When the transmission of a packet fails, a negative acknowledgement (NAK) message from the destination to the source triggers the retransmission of
the lost packets. This procedure is repeated until the packets are received successfully by the receiver. 
ARQ works well for noisy channels where the noise during
different packet transmissions are uncorrelated, and packet
errors are independent. However, in wireless communications,
packet errors are often due to channel fades, and are no
longer independent due to the correlation of the fading
process. For slow fading, or large coherence time, bursts
of packet errors may occur in consecutive transmissions.
In such cases, ARQ may not be effective and throughput performance may be degraded in the link layer. 
In recent years, cooperative methods have been successfully integrated into ARQ to overcome this problem.

Cooperative ARQ methods aim to exploit the broadcast nature of the 
wireless channel and decrease the number of retransmissions which
translates into better throughput and delay performance \cite{zv05,dlns06,yzq06,ssby06,bk11}. Broadcast property of the wireless channel
enables nodes to listen the transmitted messages from any node in their coverage area. 
When a packet transmitted from a source node
can not be decoded at the destination node, other nodes (relays) which 
have received the packet successfully
cooperate with the source and the destination at the retransmission phase. 
Cooperative ARQ aims to decrease the number of retransmissions and increase network throughput efficiency 
by using different channels which can be viewed as a special kind of spatial diversity. 
As opposed to the physical layer cooperation methods 
(see \cite{ltw04,nhh04} and the references
therein), in cooperative ARQ, relays cooperate only when the direct
link between source and destination fails.

Another way to increase network throughput is to make use of network coding 
\cite{acly00,lyn03,km03}. 
The idea behind the network coding is to combine different packets addressed 
to the same destination by performing algebraic operations. Network coding 
for wireless systems in the physical layer
has been studied extensively \cite{zll06,krwkmc08,llv10}. 
More recently, in \cite{ssm08,tnbg09,ntnb09,sv11,vtn10,fzwl09,vth11,qyzj12,mrz09,
sgjs12,avsa13},
network-coded ARQ is investigated. In \cite{ssm08,tnbg09,ntnb09,sv11}, network coding
is considered for a multicast scenario, where upon a transmission
failure, the retransmitted packet is combined with other packets
at the source.
This combination is helpful for the case when one destination
has received a damaged packet, while the other destinations have
received it correctly. Packet combining at the source can help
reduce the number of retransmitted packets, reducing queue size
\cite{ssm08}, and improving the efficiency \cite{tnbg09,ntnb09,sv11}. The methods of
\cite{ssm08,tnbg09,ntnb09} are non-cooperative network coding, since network coding
is performed by a single source node, and transmission is via
single-hop, without cooperation. In \cite{vtn10}, this approach is 
considered for the broadcast phase of a two-way relay network without
a direct link between the sources.

For cooperative networks, network coded ARQ can be implemented
where the combining of packets can be done at the relays as well.
This approach, called cooperative network-coded ARQ (C-NC-ARQ),
is promising  since it combines the diversity advantages of cooperation
with the throughput increase advantages of network coding 
\cite{fzwl09,vth11,qyzj12,mrz09,sgjs12,avsa13}.
In \cite{fzwl09}, the single-source single-hop multicast scenario of 
\cite{ssm08,tnbg09,ntnb09,sv11} is 
generalized to the case where the single source is aided by a relay.
The generalization of this idea to multiple sources is investigated in \cite{vth11,qyzj12}.
In \cite{mrz09}, authors propose a strategy where relays can combine their
own packets addressed to the destination with the retransmited packet they
are relaying. When a damaged packet is received at the destination, the relay
can combine the original packet and its own packet and transmit it to the
destination. This scenario requires the destination to have the
ability to recover network coded packets from partly damaged packets.
Similar methods are combined with physical layer two-way relay network
coding ideas in \cite{sgjs12}. C-NC-ARQ ideas were investigated within the context
of random access channels in \cite{avsa13}.

In this work, we investigate C-NC-ARQ for a two-way relay network with direct a
link between the sources. Contrary to previous work, the operation of the
relay node is more adaptive in the retransmission phase. In the proposed method,
the relay and the sources
act depending on which packets are received successfully by which nodes.
Moreover, since the correlation of the channel process is a key factor in
the performance of C-NC-ARQ methods, we investigate the performance by utilizing 
a channel model which takes
the correlation of channel errors into account. The channel with correlation
is modeled by a binary Markov process. We consider different retransmission
strategies and investigate the effect of channel correlation on the performance
of these strategies. Throughput efficiency is considered as the performance metric.
The throughput is obtained analytically, and compared with existing cooperative ARQ
methods for correlated channels. To the best of our knowledge, throughput analysis of C-NC-ARQ
for two-way relaying in correlated channel was not studied in the literature.


Outline of the rest of the
paper is as follows: In Section II, network, channel and error models are given. The proposed
C-NC-ARQ method and throughput analysis of cooperative ARQ methods
are given in Section III and Section IV, respectively. In Section V, analytical and numerical
results related to the throughput comparison of different methods are given.
Concluding remarks are given in Section VI.
\section{System Model}\label{bolum:sistem}
\begin{figure}
\centering
\includegraphics[scale=1.2]{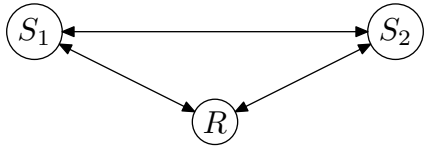}
\caption{Two-way relaying: A cooperative network consists of two sources and a relay.}
\label{sekil:network_model}
\end{figure}

There are two source nodes and a relay node in a two-way relay wireless communication
network (Fig.~\ref{sekil:network_model}). A channel between a transmitter 
receiver pair is shown by an arrow.
The nodes operate in a half-duplex mode, and the medium is time-shared, where time is
divided into slots. We assume channel reciprocity for communication in opposite directions
on a link. The channels are assumed to be flat fading and constant for a slot, but varying
between slots. 

The sources $S_1$ and $S_2$ are communicating in two ways, where $S_2$ is the destination
for $S_1$ and vica versa. Communication starts with 
$S_1$ and $S_2$ transmitting their packets
in two consecutive slots. Upon transmission of a packet, immediate feedback is sent back by the 
destination, in a stop-and-wait fashion. The packet transmission and the reception of the
corresponding feedback 
constitute a slot. The feedback is in the form of positive acknowledgement 
(ACK) or negative acknowledgement (NACK). ACKs and NACKs are assumed to be reliable. 
It is worth to emphasize that the ACK/NAK feedbacks are broadcasted over the network.
Since a packet transmitted by a node is received by
the other two nodes (Fig.~\ref{sekil:network_model}), 
at the end of a slot, all three nodes are aware of whether its
transmission is successful. 
If a transmission is not successful, the packet is assumed to be lost.

A {\em round} is defined to be group of time slots that starts with sources $S_1$ and $S_2$ 
transmitting their packets and ends with the two packets being received successfully at their
respective destinations. Consider a round starting at slot $k$. At $k$th slot, $S_1$
sends length-$M$ packet $\mathbf{p}_1$. The received signals at $S_2$
and $R$ are, respectively,
\begin{equation}
\mathbf{y}_{ss,1}[k]=\mathbf{p}_1 h_{ss}[k]+\mathbf{w}_{ss}[k],
\end{equation}
\begin{equation}
\mathbf{y}_{sr1}[k]=\mathbf{p}_1 h_{sr1}[k]+\mathbf{w}_{sr1}[k].
\end{equation}
At $(k+1)$th slot $S_2$ sends length-$M$ packet $\mathbf{p}_2$. The received signals at $S_1$ and $R$ are respectively given below:
\begin{equation}
\mathbf{y}_{ss,2}[k+1]=\mathbf{p}_2 h_{ss}[k+1]+\mathbf{w}_{ss}[k+1],
\end{equation}
\begin{equation}
\mathbf{y}_{sr2}[k+1]=\mathbf{p}_2 h_{sr2}[k+1]+\mathbf{w}_{sr2}[k+1].
\end{equation}

Here $\{\mathbf{w}_{ss}[k],\mathbf{w}_{sr1}[k],\mathbf{w}_{sr2}[k]\}$ are additive Gaussian noise:
$\mathbf{w}_{ss}[k]\sim\mathcal{CN}(\mathbf{0},\sigma_{ss}^2\mathbf{I}_M)$,
$\mathbf{w}_{sr1}[k]\sim\mathcal{CN}(\mathbf{0},\sigma_{sr1}^2\mathbf{I}_M)$,
$\mathbf{w}_{sr2}[k]\sim\mathcal{CN}(\mathbf{0},\sigma_{sr2}^2\mathbf{I}_M)$.

Time-varying statistically independent channel coefficients $\{h_{ss}(k),h_{sr1}(k),h_{sr2}(k)\}$ are
assumed to be complex Gaussian distributed: $h_{ss}(k)\sim\mathcal{CN}(0,\sigma_{h,ss}^2)$,
$h_{sr1}(k)\sim\mathcal{CN}(0,\sigma_{h,sr1}^2)$,
$h_{sr2}(k)\sim\mathcal{CN}(0,\sigma_{h,sr2}^2)$. Defining $T_p=MT_s$ as the packet duration,
$T_s$ as the symbol duration and $T_c$ as the coherence time of the channel,
the fading channels are assumed to remain constant
within one packet duration (block fading, $T_p\ll T_c$) and slowly varying between consecutive transmissions. Since
packet errors may occur in consecutive transmissions in slowly changing (highly correlated) 
block fading channels, channel correlation must be considered for ARQ because high
channel correlation may increase the number of retransmissions. Finite-state Markov models have been used to analyze the effect of
channel correlation on ARQ throughput performance \cite{wm95,zrm98,zk99_2,bl00,gil60}. 
These models assume that the channel $\{h(k)\}$ forms
a Markov chain and each $h(k)$ can be represented by finite number of states. 
When packet errors are modeled
by outages, the two-state
Markov model, which is known as Gilbert-Elliot model, is suitable \cite{gil60,eli63}.
In this model, the success/failure state of the system is directly related to the no outage/outage state of the channel.
Gilbert-Elliot model for outage
channel model is described by two channel states where the bad state $B$ represents 
the packet loss due to channel outage
and the good state $G$ represents successful transmission. The Markov chain 
has the following transition probability matrix:
\begin{equation}
\mathbf{P}=\left[ \begin{array}{cc}
P_{BB}&P_{BG}\\
P_{GB}&P_{GG}
\end{array}\right],
\label{esit:ge_model_tr}
\end{equation}
where $P_{ij}$ denotes the probability of state is $j$ at slot $k+1$, given that the
is $i$ at slot $k$. For the channels $S_1-R$, $S_2-R$, $S_1-S_2$, 
these transition probabilites are defined as 
$P_{ij,SR1}$, $P_{ij,SR2}$, $P_{ij,SS}$, and 
the state of channels at slot $k$ are represented by the variables 
$C_{SR1}(k)$, $C_{SR2}(k)$, $C_{SS}(k)$, respectively. 

In \cite{zrm97}, for complex Gaussian distributed channel process
with Jakes' spectrum, the transition probabilities are derived as:
\begin{equation}
 P_{BG}=\frac{Q(\theta,\rho\theta)-Q(\rho\theta,\theta)}{P_{ss}/(1-P_{ss})},
\end{equation}
\begin{equation}
 P_{GB}=Q(\theta,\rho\theta)-Q(\rho\theta,\theta),
\end{equation}
where $P_{ss}$ is the outage probability of the channel. The parameter $\theta$ is defined as
\begin{equation}
\theta=\sqrt{\frac{2\overline{\gamma}}{1-\rho^2}},
\end{equation}
where $\overline{\gamma}$ is average signal-to-noise ratio (SNR) of the channel and time correlation of the channel between consecutive transmissions is given by $\rho=J_0(2\pi f_m T_p)$ for
the Doppler frequency $f_m$. $Q(\cdot,\cdot)$ represents the
Marcum $Q$ function:
\begin{equation}
 Q(a,b)=\int_b^{\infty}x\exp\left(-\frac{x^2+a^2}{2}\right)I_0(ax)dx.
\end{equation}
(Similar Markov models for Rician and Nakagami flat fading channels are given in \cite{pfl04} and \cite{ks95}.)

For the direct channel between $S_1$ and $S_2$, for example, the instantaneous
SNR is $\gamma_{ss}(k) = | h_{ss}(k) | ^2 P_0/\sigma_{ss}^2$, and the average
SNR is
\begin{equation}
\overline{{\gamma}}_{ss} = \frac{ \sigma_{h,ss}^2 P_0 }{\sigma_{ss}^2}.
\end{equation}

Packet error probability can be approximated by mutual information outage probability
if strong and long channel codes are used \cite{ctb99,ml99,gc06}. 
In this case, for the desired
bit rate $R_b$ bits/symbol, the outage probability is \cite{tse:wireless}
\begin{equation}
\begin{split}
P_{ss} &= P\{ \log_2 ( 1 + \gamma_{ss}(k)) \leq R_b \} \\
&= P\{ \gamma_{ss}(k) \leq \gamma^\prime \},
\end{split}
\end{equation}
\noindent where $\gamma^\prime = 2^{R_b} - 1$ is the threshold SNR. For Rayleigh
fading channel envelope, the SNR is exponential distributed, so
\begin{equation}
P_{ss} = 1 - \exp \left( -\frac{1}{F_{s}} \right).
\end{equation}
The fading margin $F_{s}$ is defined as
\begin{equation}
F_{s} = \frac{ \bar{\gamma}_{ss} }{\gamma^\prime},
\end{equation}
which is the amount of the channel is allowed to fade below its mean value 
before an outage occurs.

For the relay channels $R-S_1$ and $R-S_2$, given the fading
margins $F_{r,1}$ and $F_{r,2}$, the error probability parameters $P_{sr,1}$ and 
$P_{sr,2}$ are similarly obtained.
\section{C-NC-ARQ Method}\label{bolum:c-nc-arq}
At the start of a round, the system is said to be in {\em transmission
phase}. The transmission phase takes two slots: in the first slot
$S_1$ transmits $\mathbf{p}_1$, and in the second slot
$S_2$ transmits  $\mathbf{p}_2$ for the first time. Unless
the direct channel $(S_1-S_2)$ is in outage in any of
these two slots, the round is completed and the next round starts,
again in transmission phase. If, on the other hand, any of
two packets is not delivered successfully at the end of
the transmission phase, the network enters {\em retransmission phase}.
What is transmitted in this phase is determined by the C-NC-ARQ table.
The retransmission phase continues until both packets are successfully
decoded by the source nodes. At the end of retransmission phase the
next round starts, in transmission phase. In retransmission phase,
if the relay has successfully received one or both packets, it cooperates
with the source nodes and retransmits the individual or network coded
packets, based on the strategy. Note that a cooperation strategy
is represented by the C-NC-ARQ table.

We assume no central control over the nodes for coordination
of signaling. The distributed coordination is achieved by reliable
ACK/NAK feedback, as a result of which every node is aware whether
a transmission is successful for the two receiving nodes at the 
end of the slot. The success/failure of the transmissions determines
the ARQ state of the network, which in turn determines the next
transmission.  The operation of the network is governed by a 
\emph{C-NC-ARQ table} which decides which packet will be transmitted by
which node in the next slot. Each row of this table corresponds
to an ARQ state of the network. All nodes in the network have this table.
The nodes listen to the broadcasted ACK/NAK feedback, keep track
of the network ARQ state, and act accordingly, without the need of
a central controller. 

Let us next explain the state model of the network. There are two types
of state variables: channel state variables, and ARQ state variables.
The channel state, denoted by $\mathbf{cs}[k]$, represents whether the channels
are in outage during slot $k$. The vector variable $\mathbf{cs}[k]$ has three elements
corresponding to one direct and two-relay channels:
\begin{equation}
\mathbf{cs}[k] = 
\begin{bmatrix}
cs(k,1) & cs(k,2) & cs(k,3)
\end{bmatrix},
\end{equation}
where $cs(k,1)$, $cs(k,2)$, $cs(k,3)$ represent the channels
$(S_1-R)$, $(S_2-R)$, $(S_1-S_2)$, during slot $k$, respectively.
The elements of $\mathbf{cs}[k]$ can take values in $\{0,1\}$:
$cs(k,i)=0$ shows that corresponding channel is in outage,
and $cs(k,i)=1$ means no outage. The ARQ state variables are
$\mathbf{ps}[k]$ and $\mathbf{rs}[k]$ 
represent the state of the packets at the end of 
$(k-1)$th slot, and their elements also take values in $\{0,1\}$. 
The vector variable $\mathbf{ps}[k]$
denotes the success/fail state of the packets $\mathbf{p}_1$ and 
$\mathbf{p}_2$ at their destinations, respectively:
\begin{equation}
\mathbf{ps}[k]=
\begin{bmatrix}
ps(k,1)&ps(k,2)
\end{bmatrix},
\end{equation}
where $ps(k,i)=0$ means the packet $\mathbf{p}_i$ is not successfully 
decoded by the other source node at the end of ($k-1$)th slot
and $ps(k,i)=1$ represents the successful decoding. 
Similarly, the vector variable $\mathbf{rs}[k]$ denotes the success/fail state of 
the packets $\mathbf{p}_1$ 
and $\mathbf{p}_2$
at the relay at the end of ($k-1$)th slot:
\begin{equation}
\mathbf{rs}[k]=
\begin{bmatrix}
rs(k,1)&rs(k,2)
\end{bmatrix}.
\end{equation}

\begin{figure}
\centering
\subfigure[][]{
\label{sekil:simulation_model}
\includegraphics[scale=0.5]{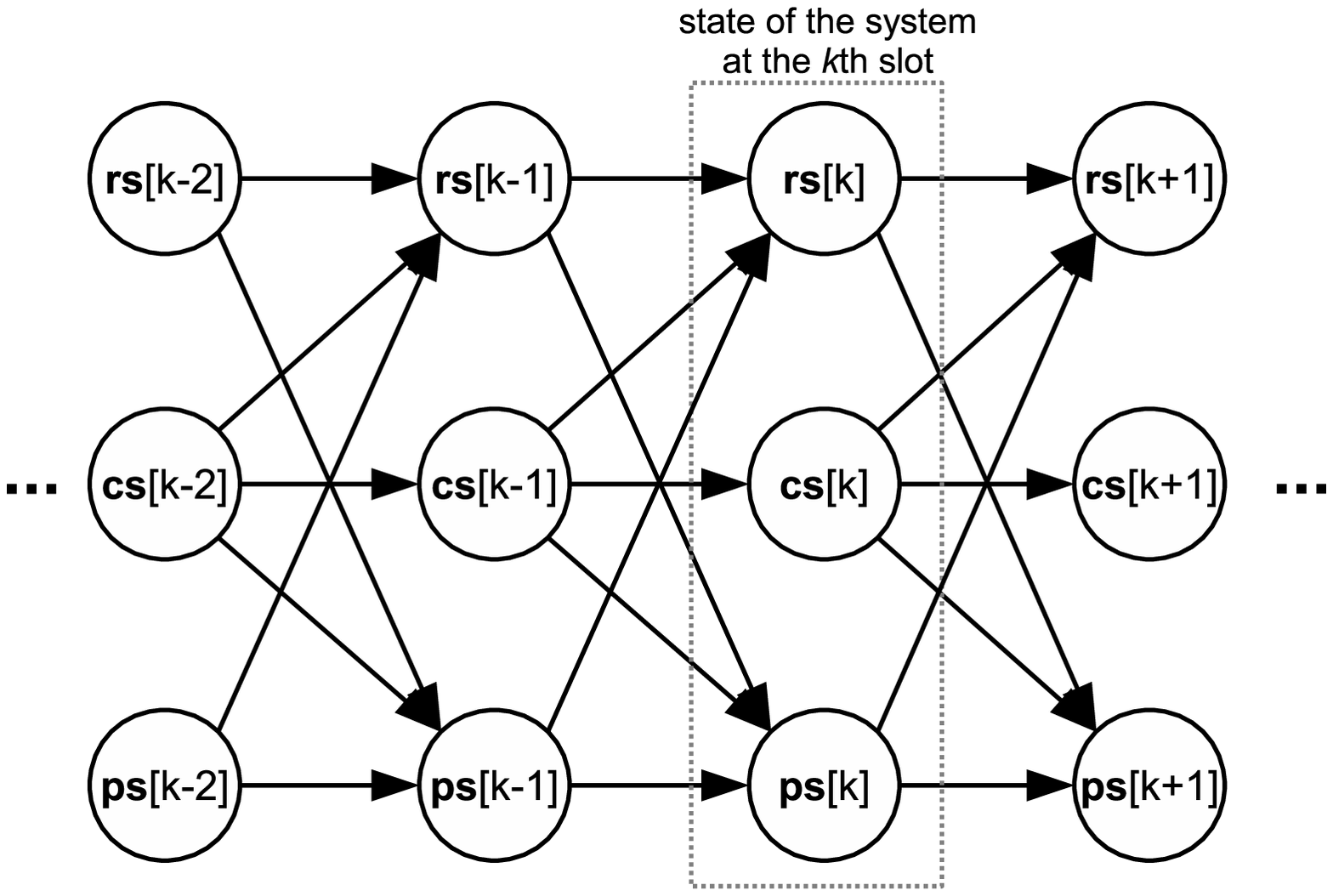}
}
\hspace{1pt}
\subfigure[][]{
\label{sekil:flow_chart_c_nc_arq}
\includegraphics[scale=0.5]{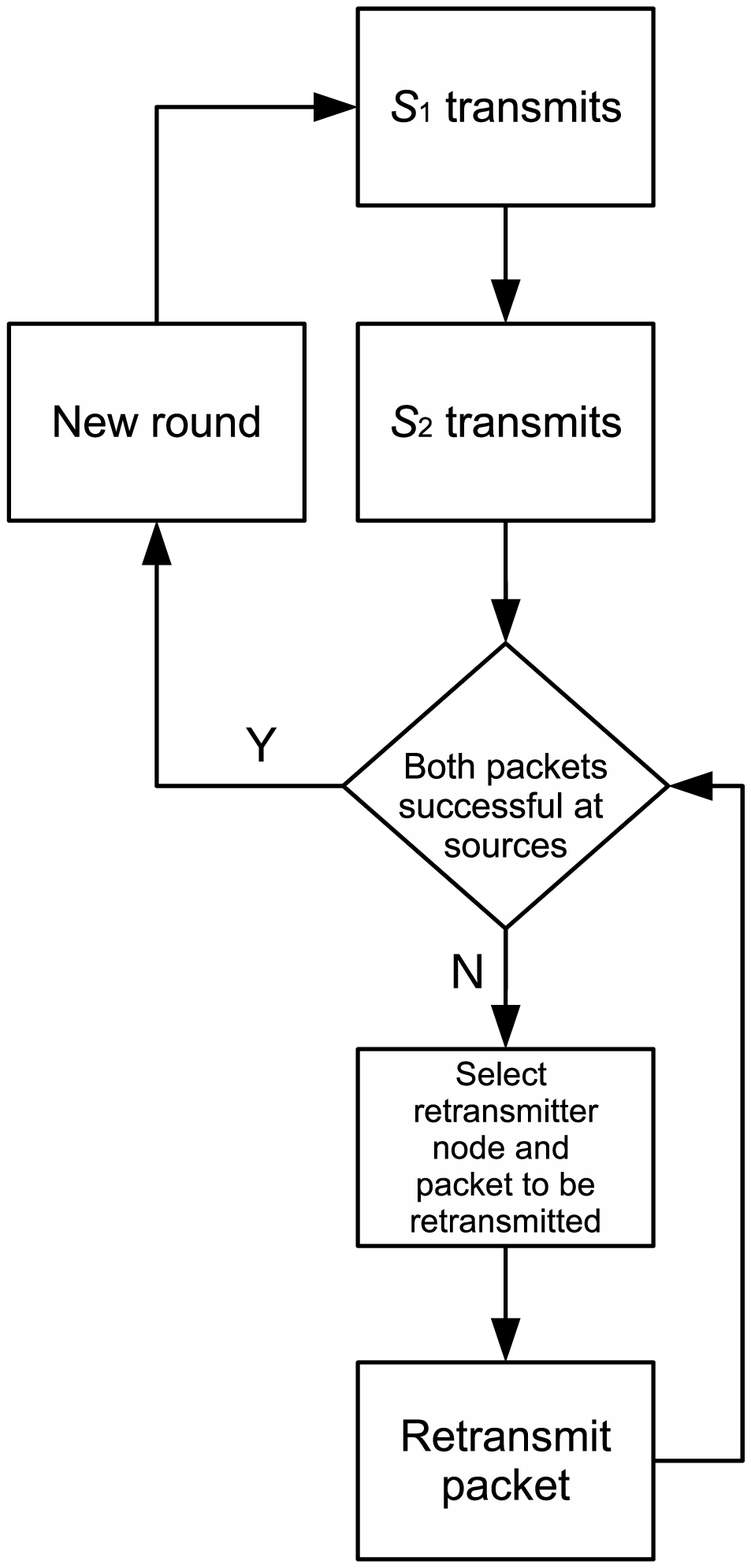}
}
\\
\caption[C-NC-ARQ state model and flow chart]
{C-NC-ARQ \subref{sekil:simulation_model} state model, \subref{sekil:flow_chart_c_nc_arq} flow chart}%
\label{sekil:c_nc_arq}%
\end{figure}

Fig.~\ref{sekil:simulation_model} depicts the dependence of the state variables over time.
In this depiction, an arrow from variable $a$ to $b$ signifies that $b$
depends on $a$. As shown in the figure, the ARQ state variables $\mathbf{ps}[k]$
and $\mathbf{rs}[k]$ depend on their previous values $\mathbf{ps}[k-1]$, $\mathbf{rs}[k-1]$, 
and also
the previous channel state $\mathbf{cs}[k-1]$. For example, if $S_1$ transmits
$\mathbf{p}_1$ at ($k-1$)th slot, ARQ state variables related to $\mathbf{p}_2$
remain the same at the end of ($k-1$)th slot: $ps(k,2) = ps(k-1,2)$,
and $rs(k,2)=rs(k-1,2)$. The ARQ state variables related to $\mathbf{p}_1$ alter depending
on the channel states at slot $k-1$: $ps(k,1) = cs(k-1,3)$, and $rs(k,1) = cs(k-1,1)$.
Note that when a new round starts, the ARQ variables $\mathbf{ps}$ and 
$\mathbf{rs}$ are initialized to zero. 

The flow chart of C-NC-ARQ is given in Fig.~\ref{sekil:flow_chart_c_nc_arq}. 
As shown by this chart, the completion of a round
depends on the condition that both packets from two source nodes are successfully decoded
by the source nodes. (Packet $\mathbf{p}_1$ from $S_1$ decoded at $S_2$, 
packet $\mathbf{p}_2$ from $S_2$ decoded at $S_1$.)
At the end of the transmission phase, if at least one packet fails, 
retransmission phase starts. 
Depending on the strategy, the retransmitting node and the retransmitted 
packet are chosen according 
to the C-NC-ARQ table which will be described in the sequel and 
the retransmissions are repeated until the
successful round condition is satisfied. 
Three new retransmission strategies are proposed: relay-based
retransmission with network coding, 
alternating retransmission with network coding, 
and channel state information based retransmission with network coding. 

All three strategies are summarized in the C-NC-ARQ table in Table~\ref{tablo:iletim_semasi}.
Each row of this table corresponds to an ARQ state shown in the left column.
On the right column the corresponding retransmission rule is shown for the
proposed methods. While the proposed methods utilize network coding,
non-network-coded versions of the methods are also shown, for comparison.
The retransmission rule is given by the notation $X \rightarrow \mathbf{p}$, which
represents the event that node $X$ transmits packet $\mathbf{p}$. For some of the
rows, the transmitting node is $C$. The node $C$ differs for the three
different strategies, the mechanisms of which will be explained in the following.

\setlength{\tabcolsep}{2pt}
\begin{table}[ht]
\centering
\begin{tabular}{| c c c c | c c |}
\hline
\multicolumn{4}{|c|}{state at the beginning $k$th slot}&\multicolumn{2}{c|}{$k$th slot}\\[1ex]
$ps(k,1)$ & $ps(k,2)$ & $rs(k,1)$ & $rs(k,2)$ & with NC & without NC\\ [1ex]
\hline
$0$ & $0$ & $0$ & $0$ & $S_1 \rightarrow \mathbf{p}_1$ & $S_1 \rightarrow \mathbf{p}_1$\\[1ex]
$0$ & $0$ & $0$ & $1$ & $S_1 \rightarrow \mathbf{p}_1$ & $S_1 \rightarrow \mathbf{p}_1$\\[1ex]
$0$ & $0$ & $1$ & $0$ & $C \rightarrow \mathbf{p}_1$ & $C \rightarrow \mathbf{p}_1$\\[1ex]
$0$ & $0$ & $1$ & $1$ & $R \rightarrow \mathbf{p}_1\oplus \mathbf{p}_2$ & $C \rightarrow \mathbf{p}_1$\\[1ex]
$0$ & $1$ & $0$ & $0$ & $S_1 \rightarrow \mathbf{p}_1$ & $S_1 \rightarrow \mathbf{p}_1$\\[1ex]
$0$ & $1$ & $0$ & $1$ & $S_1 \rightarrow \mathbf{p}_1$ & $S_1 \rightarrow \mathbf{p}_1$\\[1ex]
$0$ & $1$ & $1$ & $0$ & $C \rightarrow \mathbf{p}_1$ & $C \rightarrow \mathbf{p}_1$\\[1ex]
$0$ & $1$ & $1$ & $1$ & $C \rightarrow \mathbf{p}_1$ & $C \rightarrow \mathbf{p}_1$\\[1ex]
$1$ & $0$ & $0$ & $0$ & $S_2 \rightarrow \mathbf{p}_2$ & $S_2 \rightarrow \mathbf{p}_2$\\[1ex]
$1$ & $0$ & $0$ & $1$ & $C \rightarrow \mathbf{p}_2$ & $C \rightarrow \mathbf{p}_2$\\[1ex]
$1$ & $0$ & $1$ & $0$ & $S_2 \rightarrow \mathbf{p}_2$ & $S_2 \rightarrow \mathbf{p}_2$\\[1ex]
$1$ & $0$ & $1$ & $1$ & $C \rightarrow \mathbf{p}_2$ & $C \rightarrow \mathbf{p}_2$\\[1ex]
\hline
\end{tabular}

\caption{Cooperative (RR, AR, CR) and cooperative network coded (RR-NC, AR-NC, CR-NC) retransmission strategies.
 Retransmitting node $C$ is selected according to the strategy. $X\rightarrow \mathbf{p}$
 represents the event that the node $X$ is transmitting the packet $\mathbf{p}$.}
\label{tablo:iletim_semasi}
\end{table}

\subsection{Relay-Based Retransmission with Network Coding Strategy (RR-NC)}\label{bolum:c-nc-arq-rr}
According to the relay-based strategy, retransmissions are always executed 
by the relay if the relay
has successfully received the packets to be retransmitted. 
Thus, $C=R$ in Table~\ref{tablo:iletim_semasi}. 
If the relay does not have the packets
to be retransmitted, the retransmission is done by the original source node.

Note that network coding reduces the number of retransmissions by combining
two unsuccessful packets for the case of 
$\mathbf{rs}[k]=[\begin{array}{cc}1&1\end{array}]$, 
while the non-network coded method
RR retransmits two packets at two individual slots. 
The relay-based strategy is expected to
outperform when the fading margin of the relay channels are 
larger than the the fading margin
of the direct channel.

\subsection{Alternating Retransmission with Network Coding Strategy (AR-NC)}\label{bolum:c-nc-arq-sr}
The difference between AR-NC and RR-NC is that 
for repeated transmissions, the choice of retransmitting node alternates 
between the relay and the source node
($S_1$ or $S_2$). As an example from Table \ref{tablo:iletim_semasi}, 
when the case $\mathbf{ps}[k]=[\begin{array}{cc}0&0\end{array}]$
and $\mathbf{rs}[k]=[\begin{array}{cc}1&0\end{array}]$ occurs at the end of slot $k-1$, 
retransmission is performed by the relay ($C=R$) at the $k$th slot. 
If packet state does not
change at the end of $k$th slot ($\mathbf{ps}[k+1]=[\begin{array}{cc}0&0\end{array}]$ and
$\mathbf{rs}[k+1]=[\begin{array}{cc}1&0\end{array}]$), 
the source node is chosen as
retransmitting node ($C=S_1$) for ($k+1$)th slot. The alternating between 
the relay and the source continues for the rest
of the retransmissions. The AR-NC strategy is a little more complex
than the RR-NC strategy, since the former needs to keep track of the
last retransmitting node.
However, the AR-NC strategy is expected to improve
upon RR-NC, especially for highly correlated block fading channels where
the relay channel may enter into long duration outages.

\subsection{Channel State Information Based Retransmission with Network Coding Strategy (CR-NC)}
\label{bolum:c-nc-arq-br}
Notice that at the beginning of slot $k$, the nodes are aware of their 
channels and the channels of other nodes from the ACK/NAK feedback broadcasted 
in the previous slots. Using this information about the channel states in the 
previous slots, the choice of retransmitting node can be improved. For example, 
consider the case where $\mathbf{ps}[k]=[\begin{array}{cc}0&0\end{array}]$ and 
$\mathbf{rs}[k]=[\begin{array}{cc}1&0\end{array}]$ has occured, and the channel 
states at slot $k-1$ has been observed due to the ACK/NAK feedback. The 
retransmitting node is selected as the source ($C=S$) at the $k$th slot 
if the direct channel was not in outage while relay channel was in outage at the 
($k-1$)th slot ($cs(k-1,2)=0$, $cs(k-1,3)=1$). Otherwise the relay retransmits 
($C=R$). This strategy is expected to perform well for both slow and moderately 
fast fading since it exploits the previously observed channel states. Unless the 
channel fading very fast, the previous ACK/NAK observations will be good indicators of the 
channel states at slot $k$.

\section{Throughput Analysis}\label{bolum:is_cikarma}
Throughput analysis of the C-NC-ARQ strategies are based on the states of the network.  The three
main states of the network are defined as $T_0$ (new round state), $T_1$ (new round-2 state), and
$R$ (retransmission state). The schemes differ in how they behave when the network is in $R$ state.
The defined network state is determined by the ARQ state variables $\mathbf{ps}$, 
$\mathbf{rs}$, and also the channel state $\mathbf{cs}$. Since these state 
variables have the Markov dependence structure as shown in 
Fig.~\ref{sekil:simulation_model}, the network state also has the Markov structure in time. 

\begin{figure}
\centering
\includegraphics[scale=0.6]{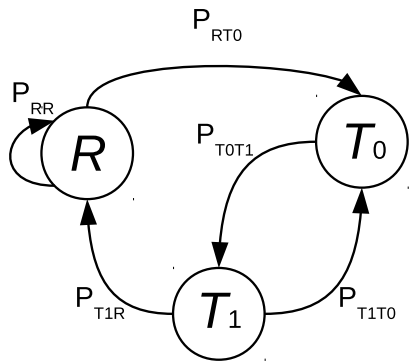}
\caption{Markov model for C-NC-ARQ.}
\label{sekil:fsmc}
\end{figure}

Let $Z(k)$ be the state of the 
C-NC-ARQ at the end of ($k-1$)th slot, 
$Z(k)=T_0$ denotes that a round has been completed successfully 
at the end of ($k-1$)th slot and a new round will start with 
$S_1$ transmitting $\mathbf{p}_1$ at $k$th slot. 
$Z(k)=T_1$ denotes that a round was completed successfully at 
($k-2$)th slot, a new round started at ($k-1$)th slot, 
$S_1$ transmitted $\mathbf{p}_1$ at ($k-1$)th slot, 
and $S_2$ will transmit $\mathbf{p}_2$ next, at $k$th slot. 
After the new round-2 state $Z(k)=T_1$, the system will either enter 
new round state $Z(k+1)=T_0$ if both packets are successful or enter 
retransmission state $Z(k+1)=R$ if at least one packet fails:

\begin{equation}
 Z(k)=\left\{\begin{array}{cc}
              T_0,&\mathbf{ps}[k]= [\begin{array}{cc}1&1\end{array}],\\
              T_1,&\mathbf{ps}[k-1]= [\begin{array}{cc}1&1\end{array}],\\
              R,&\textrm{otherwise}.
             \end{array}
             \right.
\end{equation}


The transition of $Z(k)$ states are shown in Fig.~\ref{sekil:fsmc} where 
$P_{AB}$ represents the transition
probability from state $A$ to state $B$. For the finite- state Markov model in
Fig.~\ref{sekil:fsmc} with the states $T_0$, $T_1$ and $R$, state transition 
probability matrix is given below:
\begin{equation}
\mathbf{\Sigma}=\left[\begin{array}{ccc}
  0&1&0\\
  P_{T1T0}&0&P_{T1R}\\
  P_{RT0}&0&P_{RR}
 \end{array}\right]
\end{equation}
and steady-state probabilities are $\mathbf{\pi}=\left[\begin{array}{ccc}\pi_{T,0}& \pi_{T,1}& \pi_R\end{array}\right]$ calculated from the
equation below:
\begin{equation}
 \mathbf{\pi}\mathbf{\Sigma}=\mathbf{\pi}.
\end{equation}
Since the state $T_0$ represents the new round state, whenever the system is in the state $T_0$,
it means that the packets $\mathbf{p}_1$ and $\mathbf{p}_2$ are successfully received by $S_2$ and
$S_1$, respectively. In steady-state, the ratio of expected number of 
successfully decoded packets to the total number of transmissions, 
which is defined as the average throughput, is equal to the
steady-state probability of the state $T_0$. Thus, average throughput is
\begin{equation}
 \eta=2\mathbf{\pi}_{T,0}.\label{esit:tput2}
\end{equation}
We require the elements of the transition matrix $\mathbf{\Sigma}$ to calculate the throughput. The elements of $\mathbf{\Sigma}$ may
differ depending on the strategies which are described in previous section.

In this contribution, our throughput analysis is based on the method in \cite{dlns06}. 
The variable $Z(k)$, which represents the state
of C-NC-ARQ, switches between the states $R$, $T_0$ and $T_1$ depending 
on the packet state variables $\mathbf{ps}$ and $\mathbf{rs}$, and
channel state variable $\mathbf{cs}$ according to the 
Table \ref{tablo:iletim_semasi}.

The state model of the C-NC-ARQ in Fig.~\ref{sekil:fsmc} is helpful for 
representing the main operation of the system. 
However this model is a coarse representation which hides the channel 
state and ARQ state variables. All possible configurations of these 
variables are embedded in $T_0$, $T_1$, $R$ states of the model in Fig.~\ref{sekil:fsmc}. 
In order to help analyze the system, we define \emph{sub-states} of $T_0$, $T_1$, $R$, 
for different configurations of channel and ARQ state variables. The sub-states 
of the main system states $T_0$, $T_1$ and $R$ are represented
by the state vectors $\mathbf{W}_{T,0}$, $\mathbf{W}_{T,1}$ and 
$\mathbf{W}_{R}$, respectively.
All these sub-states in these vectors constitute a new Markov model, 
which will be represented by the variable $W(k)$.

The sub-states of the new round state $T_0$ are represented 
by the vector $\mathbf{W}_{T,0}$:
\begin{equation}
 \mathbf{W}_{T,0}=\left[\begin{array}{cccccc} W_{T,0}(0) & W_{T,0}(1) & \cdots & W_{T,0}(7)\end{array}\right].
\end{equation}
For slot $k$, $W(k)=W_{T,0}(i)$ means that this slot is the first transmission 
slot of a new round, and the index $i\in\{0,\cdots,7\}$ is the channel state 
index for slot $k$. This index is the decimal corresponding the the binary vector 
$[cs(k,1),cs(k,2),cs(k,3)]$. For example, for a first transmission slot $k$, 
$[cs(k,1),cs(k,2),cs(k,3)]=101$ refers to the sub-state $W(k)=W_{T,0}(5)$.

The sub-states of the new round-2 state $T_1$ are $W_{T,1}(a,j)$ for 
$a\in\{0,\cdots,3\}$ and $i\in\{0,\cdots,7\}$, where the index $a$ is the 
decimal corresponding to $[ps(k,1),rs(k,1)]$, and $i$ is the channel state index 
for slot $k$. The reason why $[ps(k,1),rs(k,1)]$ need to be included in the 
sub-states is as follows: When the system is in state $T_0$ at the beginning of ($k-1$)th slot, 
$S_1$ transmits at the ($k-1$)th slot, so $ps(k,2)=ps(k-1,2)$ and
$rs(k,2)=rs(k-1,2)$ preserve their previous values but $ps(k,1)$ and
$rs(k,1)$ alter depending on the channel state variables $cs(k-1,3)$ and $cs(k-1,1)$, respectively:
$ps(k,1)=cs(k-1,3)$, $rs(k,1)=cs(k-1,1)$. The length-$32$ sub-state vector for $T_1$ is 
\begin{equation}
 \mathbf{W}_{T,1}=\left[\begin{array}{ccccccc} W_{T,1}(0,0) & \cdots & W_{T,1}(0,7) & \cdots & W_{T,1}(3,0) & \cdots & W_{T,1}(3,7)\end{array}\right].
\end{equation}

The sub-states of the retransmission state $R$ depend on which transmission 
strategy is used. The relay-based retransmision strategy is the simplest one 
with the least number of sub-states. We will explain the sub-states of $R$ and 
the throughput analysis for the relay-based strategy first, and later describe 
how they differ for alternating and channel state information based methods.

For relay-based retransmission strategy, the sub-states of $R$ are $W_R(b,i)$ 
for $b\in\{0,\cdots,11\}$ and $i\in\{0,\cdots,7\}$, where the index $b$ is the 
decimal corresponding to binary vector \\
$[ps(k,1),ps(k,2),rs(k,1),rs(k,2)]$ and 
$i$ is again the channel state index at slot $k$. Notice that the index $b$ has 
values not larger than $11$. This is because $ps(k,1)=ps(k,2)=1$ at the end of 
the slot $k-1$ will prompt the start of a new round at slot $k$. Let us next 
investigate the transition probabilities between the defined sub-states.

\subsection{Transition from $W_{T,0}(i)$ Sub-states}\label{bolum:gecis_to}
From state $W_{T,0}(i)$ at slot $k$, there can only be transitions to 
$W_{T,1}(a,j)$, for $i,j\in\{0,\cdots,7\}$, $a\in\{0,\cdots,3\}$. This is 
because state $T_0$ is always followed by $T_1$ in the next slot. The channel 
state $i$ at slot $k$ completely determines whether packet $\mathbf{p}_1$ 
transmitted by $S_1$ is received corectly by $S_2$ and $R$ at the end of slot 
$k$, thus it completely determines $[ps(k+1,1),rs(k+1,1)]$ and the $a$ index. Let us 
denote the $a$ index determined by the channel state index $i$ by $a^{'}$. For 
the channel state index $j$ at the next slot, all values in $\{0,\cdots,7\}$ 
are possible. Let us denote the probability of transitioning from channel state 
$i$ to channel state $j$ by $p_c(i,j)$. This probability is simply the product 
of the corresponding channels' transitions, since $S_1-R$, $S_2-R$, $S_1-S_2$ 
channels are assumed to be independent. For example, transition from channel 
state $i=2$ to $j=7$ is
\begin{align}
 p_c&(2,7)\nonumber\\
 &=P\{[cs(k,1),cs(k,2),cs(k,3)]=[010]\rightarrow [cs(k+1,1),cs(k+1,2),cs(k+1,3)]=[111]\}\nonumber\\
 &=P\{C_{SR1}(k+1)=G|C_{SR1}(k)=B\}
    P\{C_{SR2}(k+1)=G|C_{SR2}(k)=G\}\nonumber\\
    &\qquad\qquad\times P\{C_{SS}(k+1)=G|C_{SS}(k)=B\}\nonumber\\
 &= P_{BG,SR1}P_{GG,SR2}P_{BG,SS}
\end{align}
where $P_{BG,SR1}$, $P_{GG,SR2}$ and $P_{BG,SS}$ were defined in Section~\ref{bolum:sistem}.

As a result, the transition probability from $W_{T,0}(i)$ to $W_{T,1}(a,j)$ is
\begin{equation}
 P\{W_{T,0}(i)\rightarrow W_{T,1}(a,j)\}=\left\{\begin{array}{cc}
                                                 p_c(i,j)&\textrm{if }a=a^{'},\\
                                                 0&\textrm{else}.
                                                \end{array}
\right.
\end{equation}

\subsection{Transition from $W_{T,1}(a,i)$ Sub-states}\label{bolum:gecis_t1}
From $W(k)=W_{T,1}(a,i)$, there can be transitions to new round sub-states 
in $\mathbf{W}_{T,0}$ or 
retransmission sub-states $\mathbf{W}_{R}$. A transition to $W(k+1)=W_{T,0}(j)$ indicates that the 
packets $\mathbf{p}_1$ and $\mathbf{p}_2$ were received successfully in the first attempt, 
without the assistance of retransmission, and a new round starts at slot $k+1$. The 
indices $a$ and $i$, (the states $ps(k,1)$, $rs(k,1)$ and the channel state at slot $k$) 
determine whether the next state is $W(k+1)=W_{T,0}(j)$. The next state is $W_{T,0}(j)$ 
only if the following new round condition is met:
\begin{displaymath}
 a: ps(k,1)=1\qquad \textrm{and}\qquad i:cs(k,3)=1
\end{displaymath}
where the notation $a:ps(k,1)=1$ reads ``$a$ is such that $ps(k,1)=1$''. Thus the transition 
probability is 
\begin{equation}
 P\{W_{T,1}(a,i)\rightarrow W_{T,0}(j)\}=\left\{\begin{array}{cc}
p_c(i,j)&\textrm{if } a: ps(k,1)=1\quad \textrm{and}\quad i:cs(k,3)=1,\\
0&\textrm{else}.
\end{array}
\right.
\end{equation}
If the new round condition is not met, then the network enters a retransmission sub-state 
in the next slot: $W(k+1)=W_R(b^{'},j)$, where $b^{'}$is the decimal value of
\begin{equation}
\left[\mathbf{ps}[k+1], \mathbf{rs}[k+1]\right]
=\left[ps(k,1),cs(k,3),rs(k,1),cs(k,2)\right].\label{esit:wt12wr_1}
\end{equation}
To simplify, we define the notation
\begin{equation}
 b^{'}=\textrm{dec}
 \left\{\left[\mathbf{ps}[k+1], \mathbf{rs}[k+1]\right]|W(k)=W_{T,1}(a,i)\right\}.\label{esit:wt12wr_2}
\end{equation}
The notation in (\ref{esit:wt12wr_2}) tells that, given state at slot $k$ is $W_{T,1}(a,i)$, we 
know the state $ps(k,1)$, $rs(k,1)$ and the channel state at $k$, from which we can find 
$\mathbf{ps}[k+1]$ $\mathbf{rs}[k+1]$ using (\ref{esit:wt12wr_1}), and the decimal conversion gives 
$b^{'}$. Eq. (\ref{esit:wt12wr_1}) signifies that at the start of slot $k+1$, packet and relay states 
for $\mathbf{p}_1$ is the same as those at the start of slot $k$, since $\mathbf{p}_2$ was 
transmitted by $S_2$ at slot $k$. Packet and relay states for $\mathbf{p}_2$ are determined by the 
states of the channels $S_1-S_2$ and $S_2-R$, respectively. So the transition probability is 
\begin{equation}
 P\{W_{T,1}(a,i)\rightarrow W_{R}(b,j)\}=\left\{\begin{array}{cc}
                                                 p_c(i,j)&\textrm{if }b=b^{'},\\
                                                 0&\textrm{else}.
                                                \end{array}
\right.
\end{equation}

\subsection{Transition from $W_R(b,i)$ Sub-states}\label{bolum:gecis_r}
It is possible to have $W_R(b,i)\rightarrow W_R(c,j)$ or $W_R(b,i)\rightarrow W_{T,0}(j)$ 
transitions. For $W(k)=W_{R}(b,i)$, as opposed to the transitions from sub-states of $T_0$ and $T_1$, 
the transmission at slot $k$ is not fixed but it depends on $b$. For a given $b$, the states 
$\mathbf{ps}[k]$ and $\mathbf{rs}[k]$ are given, which determine what will be transmitted at slot $k$ 
using the rule in Table~\ref{tablo:iletim_semasi}. Given the transmission rule and the 
channel state at $k$, the next states $\mathbf{ps}[k+1]$ and $\mathbf{rs}[k+1]$ are found. If 
$\mathbf{ps}[k+1]=[\begin{array}{cc}1&1\end{array}]$, then $W(k+1)=W_{T,0}(j)$ with probability 
$p_c(i,j)$, the system enters a new round, and $\mathbf{ps}[k+1]$ is reset to zero. If 
$\mathbf{ps}[k+1]\neq[\begin{array}{cc}1&1\end{array}]$, then $W(k+1)=W_R(c,j)$ with probability 
$p_c(i,j)$, where $c=\textrm{dec}\{[\mathbf{ps}[k] \mathbf{rs}[k]]\}$. As an example, we provide 
the list of transitions from $W_R(b,i)$ for $b=3$ and $i\in \{0,\cdots,7\}$ in 
Table~\ref{tablo:gecis_tablosu}. The index $b=3$ corresponds to 
$\mathbf{ps}[k]=[\begin{array}{cc}0&0\end{array}]$, 
$\mathbf{rs}[k]=[\begin{array}{cc}1&1\end{array}]$, and from Table~\ref{tablo:iletim_semasi}, 
we know that $R\rightarrow \mathbf{p}_1\oplus \mathbf{p}_2$ transmission will occur at slot $k$ for 
RR-NC strategy.

\setlength{\tabcolsep}{2pt}
\begin{table}[ht]
\centering
\begin{tabular}{| c | c | c | c | c |}
\hline
$W(k)$ & \shortstack{Channel at slot $k$ \\ $S_1-R,S_2-R,S_1-S_2$}  & $\mathbf{ps}[k+1],\mathbf{rs}[k+1]$ 
& \shortstack{$W(k+1)$ \\ for $j\in\{0,\cdots,7\}$}  & \shortstack{Transition\\ probability} \\ [1ex]
\hline
$W_R(3,0)$ & $0,0,0$ & $[0\;0],[1\;1]$ & $W_R(3,j)$ & $p_c(0,j)$ \\ [1ex]
$W_R(3,1)$ & $0,0,1$ & $[0\;0],[1\;1]$ & $W_R(3,j)$ & $p_c(1,j)$ \\ [1ex]
$W_R(3,2)$ & $0,1,0$ & $[0\;1],[1\;1]$ & $W_R(7,j)$ & $p_c(2,j)$ \\ [1ex]
$W_R(3,3)$ & $0,1,1$ & $[0\;1],[1\;1]$ & $W_R(7,j)$ & $p_c(3,j)$ \\ [1ex]
$W_R(3,4)$ & $1,0,0$ & $[1\;0],[1\;1]$ & $W_R(11,j)$ & $p_c(4,j)$ \\ [1ex]
$W_R(3,5)$ & $1,0,1$ & $[1\;0],[1\;1]$ & $W_R(11,j)$ & $p_c(5,j)$ \\ [1ex]
$W_R(3,6)$ & $1,1,0$ & $[1\;1],[1\;1]$ & $W_{T,0}(j)$ & $p_c(6,j)$ \\ [1ex]
$W_R(3,7)$ & $1,1,1$ & $[1\;1],[1\;1]$ & $W_{T,0}(j)$ & $p_c(7,j)$ \\ [1ex]
\hline
\end{tabular}
\caption
{Transitions from $W_R(3,\cdot)$ for RR-NC strategy.}
\label{tablo:gecis_tablosu}
\end{table}

\subsection{Steady State Probabilities}
In order to obtain the average throughput in (\ref{esit:tput2}), we need the steady state 
probabilities of the sub-states defined. To find the steady state distribution, we define the 
overall sub-state vector
\begin{equation}
 \mathbf{W}_o=\left[\begin{array}{ccc} \mathbf{W}_{T,0} & \mathbf{W}_{T,1} & \mathbf{W}_{R}\end{array}\right].
\end{equation}
The length of $\mathbf{W}_o$ for the relay based retransmission strategy is $136$. Next we construct 
the overall probability transition matrix $\mathbf{P}_o$. The $(m,n)$th element of $\mathbf{P}_o$ is:
\begin{equation}
P_o(m,n)=P\{W_o(m)\rightarrow W_o(n)\}\qquad\textrm{for }m,n\in\{1,\cdots,136\}.
\end{equation}
The matrix $\mathbf{P}_o$ is constructed using the sub-state transition probabilities explained in 
Subsections \ref{bolum:gecis_to}, \ref{bolum:gecis_t1}, \ref{bolum:gecis_r}, and has the following structure:
\begin{equation}
\mathbf{P}_o=\left[\begin{array}{ccc}
  \mathbf{0}&\mathbf{P}_{o,T_0T_1}&\mathbf{0}\\
  \mathbf{P}_{o,T_1T_0}&\mathbf{0}&\mathbf{P}_{o,T_1R}\\
  \mathbf{P}_{o,RT_0}&\mathbf{0}&\mathbf{P}_{o,RR}
 \end{array}\right].
\end{equation}
The vector of steady state probabilities
\begin{equation}
 P\{\mathbf{W}_o\}=\left[\begin{array}{ccc} P\{\mathbf{W}_{T,0}\} 
 & P\{\mathbf{W}_{T,1}\} & 
 P\{\mathbf{W}_{R}\}\end{array}\right]
\end{equation}
is found from the solution of 
\begin{displaymath}
 P\{\mathbf{W}_o\}=P\{\mathbf{W}_o\}\mathbf{P}_o\quad\textrm{and}\quad\sum_{i=1}^{136}P\{W_o(i)\}=1.
\end{displaymath}
Finally, the steady state probability of $T_0$ for the throughput in (\ref{esit:tput2}) is obtained 
as
\begin{equation}
\mathbf{\pi}_{T,0}=\sum_{i=1}^8 P\{W_{o}(i)\}.
\end{equation}

\subsection{Alternating and Channel State Information Based Retransmission Strategies}
The throughput analyses for the alternating (AR-NC and AR) and channel state information based 
(CR-NC and CR) retransmission strategies are similar, except for the fact that the number of substates in 
$\mathbf{W}_R$ increases.

For AR-NC and AR, we define a token index $t\in\{0,1\}$ that alternates between $0$ and $1$ with each 
retransmission by $C$ in Table~\ref{tablo:iletim_semasi}. If $t=0$ then $C=R$ will retransmit, otherwise 
one of $S_1$ and $S_2$ will retransmit based on which packet is transmitted. The sub-states of $R$ 
are denoted by $W_R(b,i,t)$, and there are $12\times 8 \times 2=192$ sub-states in the sub-state vector 
$\mathbf{W}_R$.

Similar to alternating retransmission strategies, 
there is a token variable that controls the retransmitting node for channel state information based retransmission strategies
(CR-NC and CR). In this case, not all the sub-states include the token variable, 
just the rows in Table \ref{tablo:iletim_semasi} which includes $C$. For example,
$W_R(0,\cdot)$ does not include token variable whereas $W_R(7,\cdot,\cdot)$ does. 
According to the state $W_R(7,t,\cdot)$, retransmission is realized by
relay if $t=0$ else $S_1$ retransmits. 
Unlike the operation in AR-NC or AR, token variable $t$ is not altered after 
every transmission because retransmitting node
is selected according to the previous channel state variable $\mathbf{cs}$. 
So, the number of sub-states in $\mathbf{W}_R$ is
$5\times2\times8+7\times8=136$ for CR-NC, and $6\times2\times8+6\times8=144$ for CR.

\section{Numerical Results}\label{bolum:sayisal}
In this section we provide performance results of C-NC-ARQ methods for 
different channel conditions, and observe the effect of
channel correlation on the network throughput performance. 
The simulation results are obtained using Monte Carlo 
simulations where fading channels are
randomly generated using (\ref{esit:ge_model_tr}) and 
the protocol rules given 
in Table \ref{tablo:iletim_semasi}.

\begin{figure}
\centering
\includegraphics[scale=0.6]{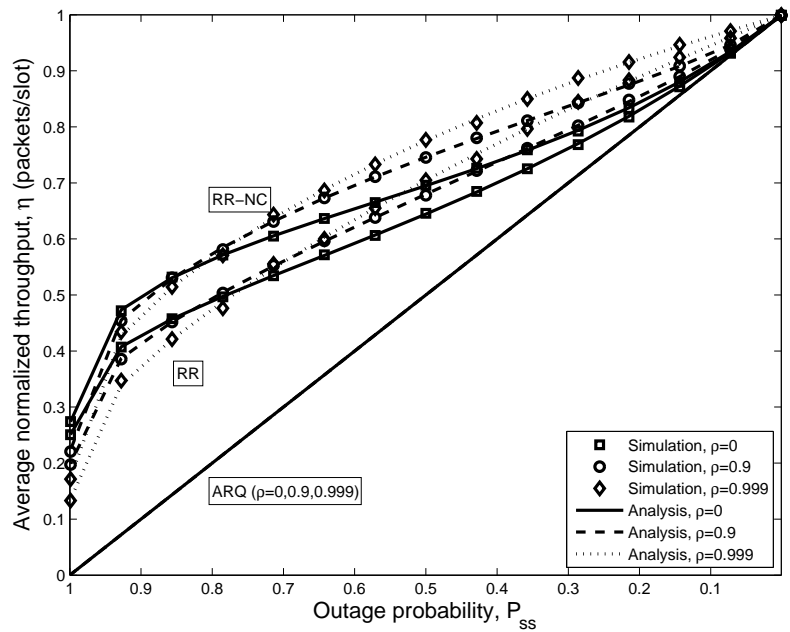}
\caption{Analytical and simulation results of relay-based retransmission strategies for different correlation coefficients. $F_r/F_s=10$ dB.}
\label{sekil:pe_vs_tput_rr}
\end{figure}

In Fig.~\ref{sekil:pe_vs_tput_rr}, network throughput 
performance of RR-NC and RR can be 
seen for different correlation coefficients. Three different correlation 
coefficients are examined: uncorrelated ($\rho=0$), 
highly correlated ($\rho=0.9$), 
fully correlated 
($\rho=0.999$) cases. Analytical results are compared with the 
Monte-Carlo simulalation 
results, and it is observed that analytical and simulation results coincide. 
The case where 
the fading margins of relay channels are higher than that of direct channel is 
considered: $F_r/F_s=10$ dB where $F_{r,i}=F_r$ for $i=1,2$. As a comparison, the throughput 
of conventional stop-and-wait ARQ is also shown, which is $\eta_{\textrm{ARQ}}=1-P_{ss}$. 
It is observed that for large values of outage probability (for $P_{ss}>0.8$), 
the channel 
correlation $\rho$ has a negative impact on the throughput performance. This is due to the 
cases where the retransmission phase is locked in repeated relay retransmission, whose 
channel is in a long-duration outage. Such 
a threshold for $P_{sr}$ can be defined as $0.15$ for the case where $F_r/F_s=10$ dB. 
For $P_{ss}<0.8$ and $P_{sr}<0.15$, we observe the positive impact of 
channel correlation. This is explained by low probability of outage combined with the 
diversity advantage of the relay mean that highly correlated block fading channels result in long-duration 
good state channels. Another important observation is that network coding can improve 
throughput by $0.1$, which is a significant improvement. 

\begin{figure}
\centering
\includegraphics[scale=0.6]{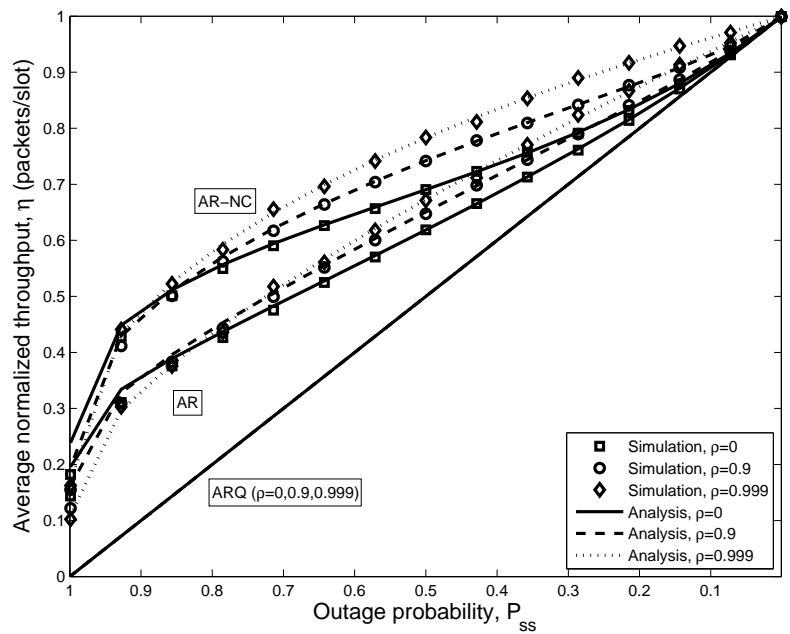}
\caption{Analytical and simulation results of alternating retransmission strategies for different correlation coefficients. $F_r/F_s=10$ dB.}
\label{sekil:pe_vs_tput_sr}
\end{figure}

\begin{figure}
\centering
\includegraphics[scale=0.6]{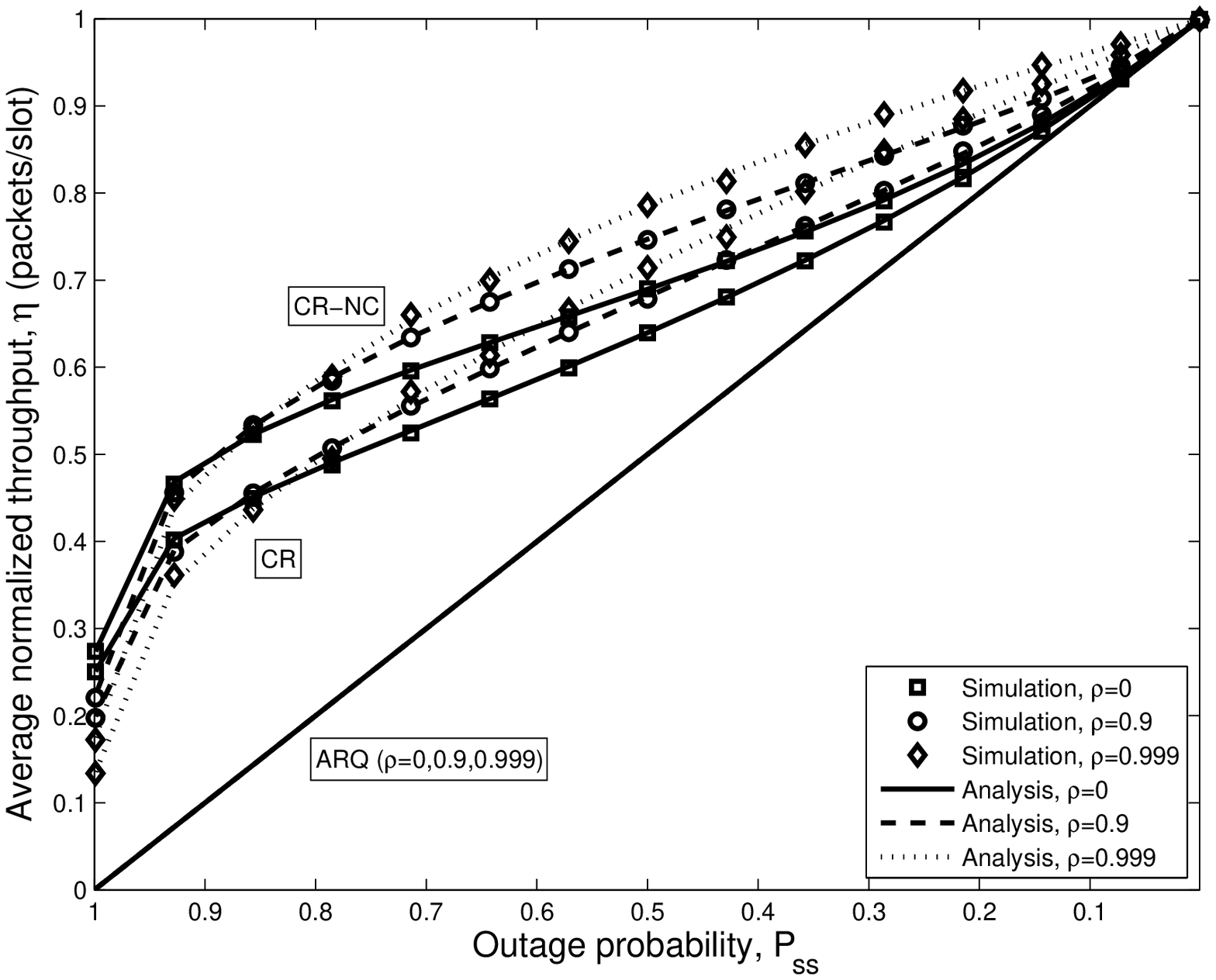}
\caption{Analytical and simulation results of channel state information based retransmission strategies for different correlation coefficients. $F_r/F_s=10$ dB.}
\label{sekil:pe_vs_tput_br}
\end{figure}

Similar behaviors are observed for the alternating retransmission strategy in 
Fig.~\ref{sekil:pe_vs_tput_sr} and channel state information based retransmission strategy 
in Fig.~\ref{sekil:pe_vs_tput_br}.

\begin{figure}
\centering
\includegraphics[scale=0.6]{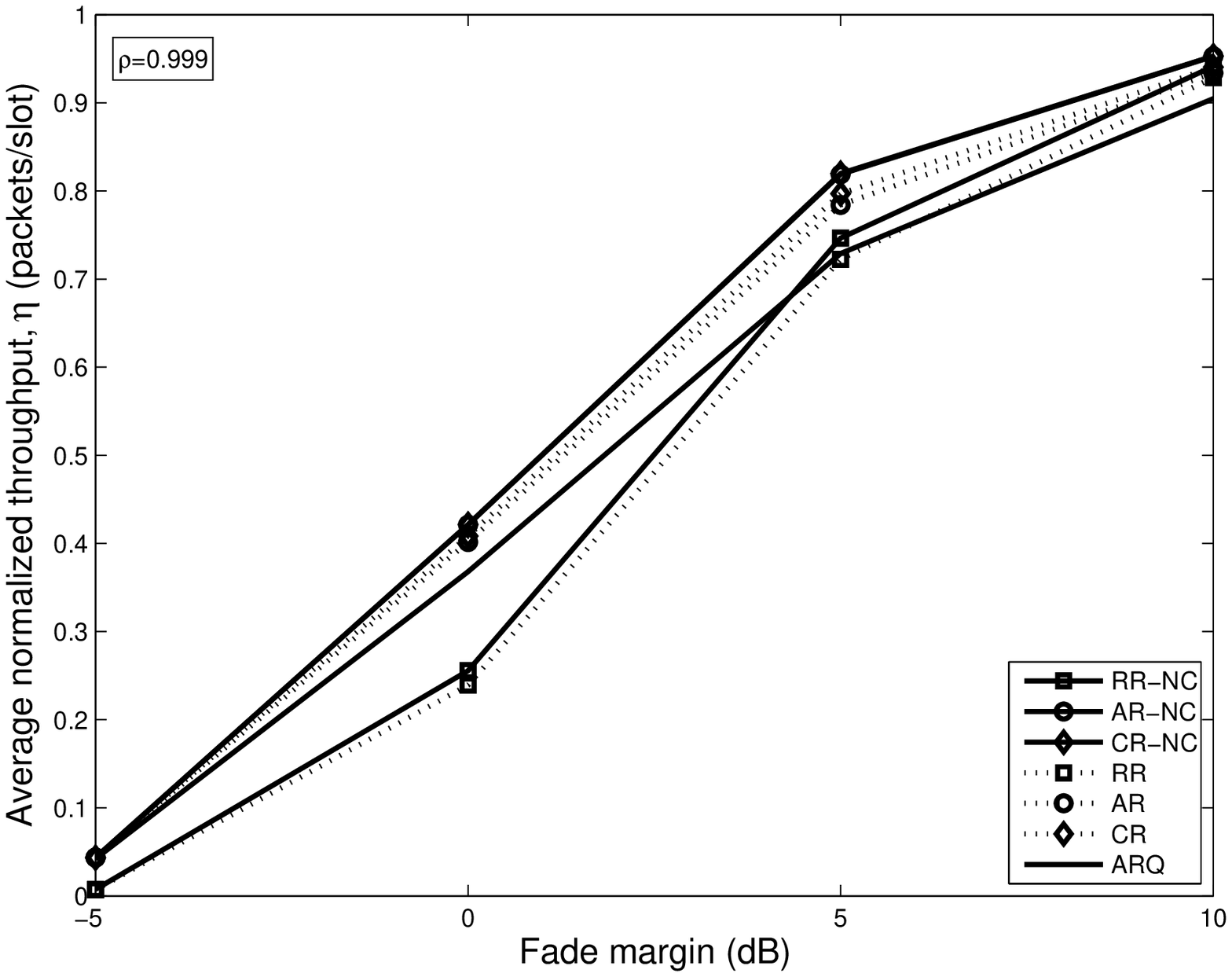}
\caption{Performance comparison of retransmission strategies as a function of fading margin. $F_r/F_s=0$ dB.}
\label{sekil:Fs_vs_tput}
\end{figure}

 The three retransmission strategies are compared in Fig.\ref{sekil:Fs_vs_tput}. In this 
 figure, only analytically obtained throughput results are shown. Fully correlated 
 case of the channel is investigated in Fig.\ref{sekil:Fs_vs_tput}, as a function of the 
 fading margin, where the relay channels have the same fading margins as the direct channel, 
 $F_r/F_s=0$ dB. For the case of high correlation, $\rho=0.999$, the relay based retransmission 
 strategy performs worse than AR, CR and even traditional ARQ, especially for low fading margin. This is 
 due to the fact that RR strategy repeatedly attempts to retransmit from consequtive bad relay 
 channels. For the AR and CR strategies, this situation does not occur.

\begin{figure}
\centering
\includegraphics[scale=0.6]{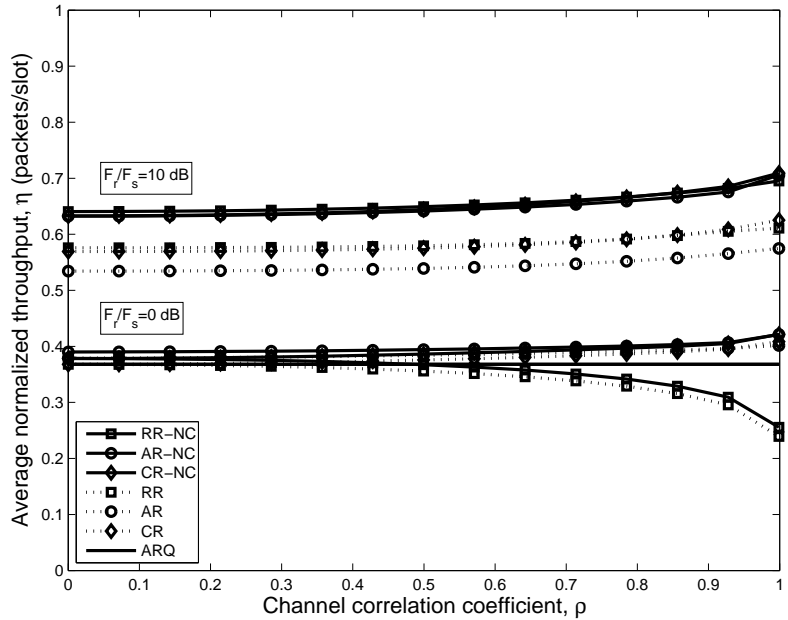}
\caption{Performance comparison of retransmission 
strategies as a function of channel correlation. $F_s=0$ dB.}
\label{sekil:rho_vs_tput}
\end{figure}

 In Fig.~\ref{sekil:rho_vs_tput}, the throughput performances are shown as a function of the correlation 
 coefficient for fixed $F_s=0$ dB and the cases where the relay channels have better fading 
 margins than the direct channel ($F_r/F_s=10$ dB), and where the relay channels have the 
 same fading margins as the direct channel. For the case when the relay channels have good 
 average reliability ($F_r=10$ dB), we observe the  gain due to the network coding for all 
 three strategies whereas for the $F_r=0$ dB case the improvement is not so significant. This 
 is due to the fact that, in order to see the network coding advantage the ARQ state 
 $\mathbf{ps}[k]=[\begin{array}{cc}0&0\end{array}]$, 
 $\mathbf{rs}[k]=[\begin{array}{cc}1&1\end{array}]$ 
 in Table~\ref{tablo:iletim_semasi} needs to occur frequently, which happens when the relay 
 channels are better than the direct channel on average. For the $F_r=10$ dB case, the relay 
 based retransmission strategy outperforms the alternating retransmission strategy because 
 the relay channels are better than the direct channel on average and alternating between 
 relay and source degrades performance for this case. It is observed that unless the channel 
 correlation is ver low, the channel state information based strategy performs best among the 
 three strategies. For channel correlation close to zero, the channel state of the previous 
 slot provides no information about the current slot so the choice of the channel state 
 information based strategy becomes almost arbitrary. For the $F_r=0$ dB case, the relay 
 based retransmission strategy performs worst  because it insists on repeated transmissions 
 from the relay eventhough the relay channel may not be in a good state for repeated 
 slots, especially for large values of channel correlation.

\begin{figure}
\centering
\includegraphics[scale=0.6]{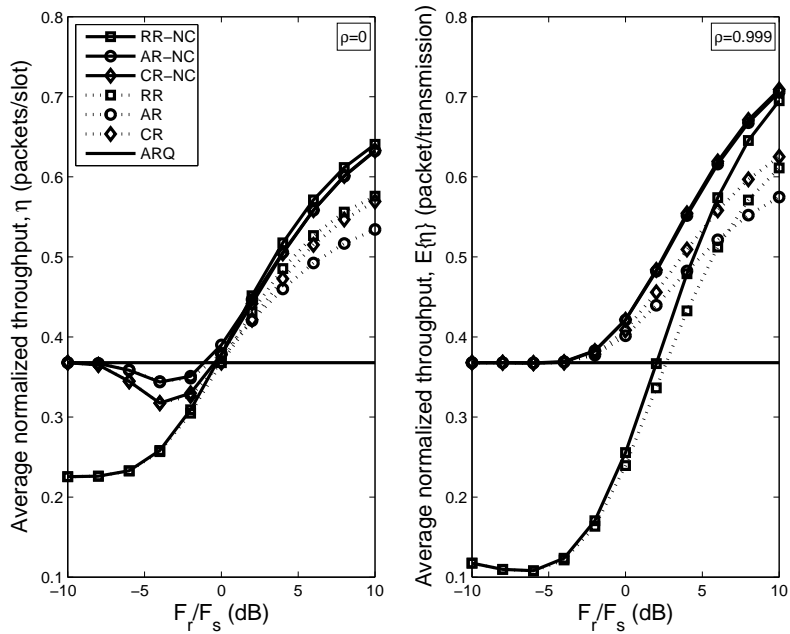}
\caption{Performance comparison of retransmission strategies 
as a function of the ratio of fading margins of direct and relay channels. $F_s=0$ dB.}
\label{sekil:scale_vs_tput}
\end{figure}

 The performances are shown as a function of the relay fading margin in Fig.~\ref{sekil:scale_vs_tput}. 
 As expected, the relay based retransmission strategy is poor for low values of relay channel 
 fading margin. For large values of $F_r$, the relay based retransmission strategy works 
 well, slightly better than the channel state information based retransmission for $\rho=0$, 
 and slightly worse than the channel state information based retransmission for $\rho=0.999$. 
 It is observed that for the case where the relay channel is worse than the direct channel 
 ($F_r/F_s<0$ dB) and very low correlation values ($\rho\ll 1$), cooperative ARQ strategies 
 may actually perform worse than the traditional ARQ. When the relay channel is much better 
 than the direct channel ($F_r/F_s>10$ dB) relay based retransmission can be a good choice, 
 otherwise alternating and channel state information based strategies work well.

\begin{figure}
\centering
\includegraphics[scale=0.6]{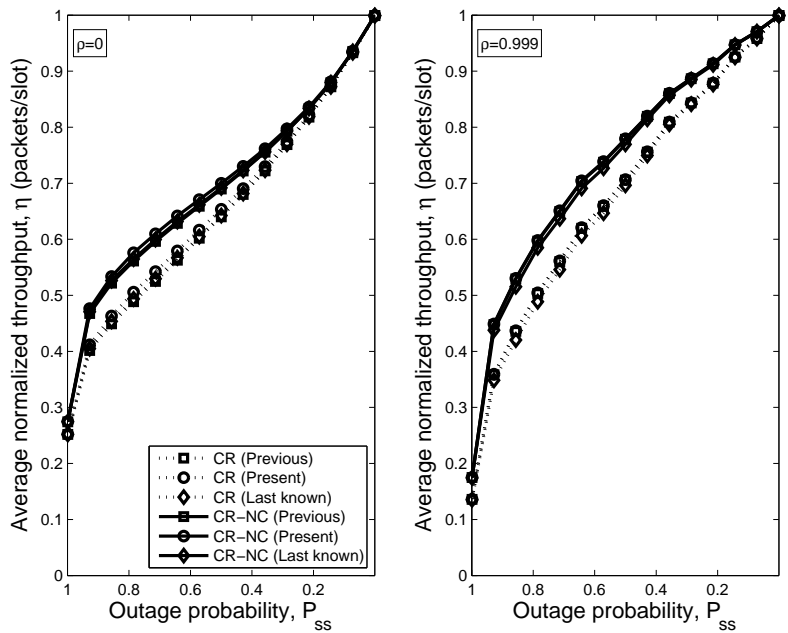}
\caption{Performance comparison of channel state information based 
retransmission strategies. $F_r/F_s=10$ dB.}
\label{sekil:pe_vs_br_all}
\end{figure}

Finally, we note that in the analysis of the channel state information based retransmission 
strategy, it was assumed that the channel state information of the previous slot 
(($k-1$)th slot) is available for deciding the transmission at slot $k$. The channel state 
information is going to be obtained utilizing the ACK/NAK feedback at each slot. However, the 
ACK/NAK feedback of all chanels may not be available for the previous slot. In practice, the 
latest received ACK/NAK feedback is going to be used as the \emph{last known} state of each 
channel, which may degrade the performance since the channel state information may be outdated. 
In order to investigate this effect, we provide Fig.~\ref{sekil:pe_vs_br_all}, where the 
channel state information based retransmission strategy using channel state information of 
previous slot and the last known slot are compared. For comparison, we also show the performance 
of the case where the channel state information of the current slot is utilized. It is observed 
that the throughput performances are very close.

\section{Conclusion}
In this paper, novel cooperative ARQ methods which integrate network coding
into retransmission phase are proposed and performance of the proposed methods are
analyzed for two-way relay network. An analytical method is derived for obtaining
network throughput for correlated channels and is utilized to compare different cooperative ARQ
methods for different channel settings. It is observed that unless the average outage rate 
of the relay channels are worse than the direct channel and the channel is very fast fading, 
the proposed strategies improve performance. The impact of network coding is seen when the 
relay channels have a fading margin of $10$ dB or larger. Channel correlation in time 
improves the gain of the proposed methods in general. Among the proposed retransmission 
strategies, relay based retransmission is the simplest one, and can be a good choice if the 
average reliability of the relay channel is good. The channel state information based 
retransmissionis the best strategy if the nodes can keep track of the last known channel 
state information.

The generalization of the methods and their analyses to a more general network model with 
more relays and sources remains as a future work.


\bibliographystyle{IEEEtran}
\bibliography{IEEEabrv,referans}


\end{document}